\documentclass[a4paper]{PoS}
\usepackage{amsmath,pstricks}
\usepackage{epsfig, enumerate}
\usepackage{algorithm}
\usepackage{algpseudocode}
 
\def\as{\alpha_s}
\def\asb{\bar{\alpha}_s}
\def\bT{\mathbf{T}}
\def\bt{\mathbf{t}}
\def\ibf{\imath\,\mathbf{f}}

\def\cW{\mathcal{W}}

\def\X{\footnotesize X}
\def\CF{\mathrm{C_F}}
\def\CA{\mathrm{C_A}}
\title{Eikonal gluon radiation at finite-N$_c$ beyond 2 loops}

\ShortTitle{Eikonal gluon radiation at finite-N$_c$}

\author{\speaker{Kamel KHELIFA-KERFA} \\
        D\'{e}partement de Physique, Facult\'{e} des Sciences\\
        Universit\'{e} Hassiba Benbouali de Chlef - Chlef, Algeria\\
        E-mail: \email{kamel.kkhelifa@gmail.com}
        }
\author{Yazid DELENDA          \\
        D\'{e}partement des Sciences de la Mati\`{e}re, Facult\'{e} des Sciences\\
        Universit\'{e} Hadj Lakhdar - Batna, Algeria\\
        E-mail: \email{yazid.delenda@gmail.com}\\
        }

\abstract{We present first calculations of QCD matrix-elements in perturbation theory at finite N$_c$ beyond 2 loops in the eikonal approximation for $e^+ e^-$ annihilation processes. For the emission of $n$ soft energy-ordered gluons we solve both the colour and kinematic structures at a given order in perturbation theory by means of a {\tt Mathematica} program that relies solely on a recently developed {\tt Mathematica} code, {\tt ColorMath}, that evaluates the trace of products of colour matrices. At large N$_c$, our squared amplitudes  reduce to those already known in the literature.
}

\FullConference{XXIII International Workshop on Deep-Inelastic Scattering\\
		 April 27 - May 1, 2015\\
		 Dallas, Texas}

\begin{document}

\section{Introduction}

The recent runs of the Large Hadron Collider (LHC) at CERN, the substantial data collected and its subsequent analyses have opened a new era of what might be referred to -in analogy with the Electroweak theory- as ``QCD precision measurements''. For instance, the strong coupling constant $\alpha_s$ has recently been determined by the CMS collaboration with experimental uncertainties that are much smaller than their theoretical counterparts \cite{Khachatryan:2014waa}. Moreover, some recent experimental analyses even concluded with a stress on the necessity for higher fixed-order calculations and/or matching with resummed results in order to better describe the data \cite{Chatrchyan:2014gia}. It is thus necessary, perhaps more than ever before, that more efforts should be spent by the theory community on ``QCD precision calculations''. For this reason, and others, the present work has been carried out.     

It is well known that the main limiting factor in QCD calculations is the scattering amplitude (or matrix-element). Typical, and generally only possible, QCD calculations rely on perturbation theory (PT) in which the matrix-element is expanded as a series in the strong coupling $\alpha_s$ which becomes small at high energies (asymptotic freedom) ensuring the convergence of the series. In most QCD processes only the first few orders in PT expansion have been computed, with the most difficult calculational challenges coming from virtual (loop) Feynman diagrams.   

One of the very useful approximations used during the last few decades whereby calculations of matrix-elements are substantially simplified -particularly virtual corrections- is the ``eikonal approximation'' (or equivalently the ``soft insertion rules'') \cite{Levy:1969cr, Abarbanel:1969ek, Wallace:1973iu}. In QCD, this approximation corresponds to the limit where the momenta of the radiated gluons are soft. The standard Feynman rules are then replaced by the effective eikonal Feynman rules. Amongst the important characteristics of the eikonal approximation is its all-orders exponentiation for both abelian and non-abelian theories \cite{Yennie:1961ad, Gatheral:1983cz, Frenkel:1984pz}. 

Nonetheless, even in the aforementioned approximation, the calculations of even the first orders in PT expansion of QCD matrix-elements at {\it finite}-N$_c$ (N$_c$ being the number of quark colours) have proven delicate and radiative corrections of up to two gluons had been the sate-of-the-art for quite a while. The chief reason for such a serious limitation in the QCD precision calculations programme  is two-fold:\footnote{We are only discussing tree-level matrix-elements. Loop corrections can be straightforwardly treated in the eikonal approximation, as shall be hinted at later.}
\begin{itemize}
 \item The matrix-valued (non-commutative) colour space of QCD.
 \item The factorial growth of the number of Feynman diagrams at each escalating PT order. 
\end{itemize}

A partial solution to the above hindrances that was employed since quite a while is that related to the large-N$_c$ limit \cite{Bassetto:1984ik, Fiorani:1988by}. In this limit, the colour space effectively becomes scalar-valued and the number of Feynman diagrams reduces substantially due to discarding non-planar diagrams (suppressed by $1/\mathrm{N}_c^2$). In fact, an analytical compact form for the emission of $n$ soft energy-ordered gluons at large-N$_c$ was reported in \cite{Bassetto:1984ik, Fiorani:1988by}, and is generally implemented in Monte Carlo parton showers (dipole cascade picture, e.g., \cite{Lonnblad:1992tz}).

A full solution, whereby one restores the full colour structure and spans over all possible planar and non-planar contributions, has not yet been properly addressed in the literature. It is the aim of this work to carry out this very task. Firstly, the problem of the non-abelian colour structure is resolved via the aid of the {\tt ColorMath} program developed by Sj\"odahl \cite{Sjodahl:2012nk}. It is a {\tt Mathematica} package that performs colour-summed calculations in SU(N$_c$) for N$_c \geq 2$. Secondly, the factorial cascade gluon branching (in a dipole-like picture for energy-ordered gluons) is fully taken account of via a {\tt Mathematica} code that we have developed under the name of {\tt EikAmp}. \footnote{The final version of the {\tt EikAmp} package has not been finalised yet and will be presented in a future publication.} The program automatically computes the eikonal amplitude squared for all possible real and/or virtual gluon configurations at finite N$_c$ at (theoretically) any given order in PT. It relies solely on {\tt ColorMath} and the built-in {\tt For} {\tt loop} procedure. 

The output of the program is however a lengthy cumbersome expression for which we provide, in ref. \cite{Delenda:Preparaion}, compact forms up to 4 loops (and partially at 5 loops). Finite-N$_c$ corrections are found to be absent at 3 loops and first appear at 4 loops. Noticeably at 4 loops the amplitude squared exhibits some characteristics that were claimed to be absent by other authors \cite{Dokshitzer:1991wu}. Moreover, finite-N$_c$ corrections to the eikonal amplitude squared seem to have some peculiar properties (and consequently the amplitude squared itself) that are absent for large-N$_c$ contributions, and on which we shall briefly shed some light in the last section.

In the next section we introduce the eikonal approximation as well as the main eikonal Feynman rules that will prove essential to our later calculations. Moreover we present, in the same section, the final form of the eikonal amplitude squared for the emission of $n$ energy-ordered soft gluons, and describe the skeleton of the {\tt EikAmp} program. After that we discuss the main important features of the all-orders form of the eikonal amplitude squared and finally summarise  in the last section.

\section{Eikonal amplitudes}

\subsection{Eikonal approximation}

\begin{figure}[h!]
 \centering
 \includegraphics[scale=.5]{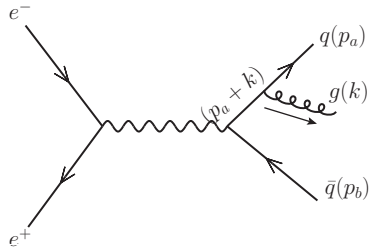}
 \caption{One of the Feynman diagrams contributing to the simple QCD process: $e^+e^- \to q + \bar{q} + g$.}
 \label{fig:SimpleQCDProcess}
\end{figure}

Consider the simplest QCD process $e^+e^- \to q \bar{q}$ accompanied with the emission of a gluon $g$ with 4-momentum $k$, as illustrated in figure \ref{fig:SimpleQCDProcess}.

The quark propagator reads:
\begin{equation}
  D_q = \frac{\imath \left(p_{a}^{\mu} \gamma_{\mu} + k^\mu \gamma_\mu \right)}{ \left(p_a + k \right)^2 }\,,
  \label{eq:FullQuarkProp}
\end{equation}
where $\gamma_\mu$ are the usual Dirac matrices. The eikonal approximation corresponds to the situation where the gluon $k$ is soft. In other words, when the momentum $k \to 0$ and thus $p_a \sim p_b$. In this case, the propagator above, eq. \eqref{eq:FullQuarkProp}, simplifies to:
\begin{equation}
 D_{q}^{\mathrm{eik}} = \frac{\imath p_a^\mu \gamma_\mu}{2 p_a \cdot k}\,,
\end{equation}
where on-mass-shell condition is assumed ($p_a^2 = 0$). Therefore one may deduce the following eikonal Feynman rules for a (anti)quark propagator:
\begin{figure}[h!]
\centering
\includegraphics[scale=.8]{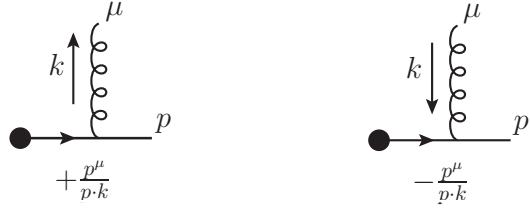}
\caption{Eikonal Feynman rules for (anti)quark propagators.}
\end{figure}

Analogous Eikonal rules (for the emission and absorption of a soft gluon) hold for gluon propagators.
Using the above rules one can build up the eikonal amplitude for any given process and/or gluon configuration, as we shall demonstrate in the next subsection. An important feature of eikonal amplitudes which is manifested even for the first emission is that they {\it factorise} into a product of a Born amplitude times the sum of all possible emitting dipole-legs \cite{Dokshitzer:1991wu, Laenen:2010uz}. This factorisation property stands at the heart of the all-orders exponentiation of eikonal amplitudes discussed in the introduction.   

It is worth mentioning that in the eikonal approximation virtual corrections amount simply to assigning a minus sign to the corresponding real emission squared amplitude, as is explicitly shown in ref. \cite{Delenda:Preparaion}. This is of course true for the softest gluon, i.e., the amplitude squared for a given configuration in which the softest gluon is virtual equals minus the squared amplitude for the same configuration but with the softest gluon being real. The general form of the eikonal amplitude squared for configurations in which an arbitrary number of virtual gluons are present is discussed in what follows below. Due to the latter equality between real and virtual contributions --up to a sign-- we shall refer to PT orders as loop orders.

\subsection{General formalism}

The emission amplitude of a (real) soft gluon $k$ by an ensemble of: a quark $p_a$, an antiquark $p_b$ and $m$ energy-ordered harder gluons $k_i$ is given by \cite{Dokshitzer:1991wu, Forshaw:2008cq}:
\begin{equation}
 \mathcal{M}(p_a, p_b, k_1, \hdots, k_{m}, k) = g_s \left[ \frac{p_a \cdot \epsilon_a^{\ast c}}{p_a \cdot k } \bT_a^c +  \frac{p_b \cdot \epsilon_b^{\ast c}}{p_b \cdot k } \bT_b^c + \sum_{i=1}^m \frac{k_i \cdot \epsilon_i^{\ast c}}{k_i \cdot k } \bT_i^c\; \right]  \mathcal{M}(p_a, p_b, k_1, \hdots, k_{m})\,,
 \label{eq:EikAmp_A}
\end{equation}
where $\bT_{i,\, i=a,b,1,\hdots,m}^c$ is the colour operator represented by $+(-) \bt^c_i$ (fundamental representation) if the emitting parton $i$ is an outgoing (incoming) quark or incoming (outgoing) antiquark, and $-(+) \ibf^c_i$ (adjoint representation) if the emitting parton is an outgoing (incoming) gluon \cite{Forshaw:2008cq}. The term $\epsilon_i^{\ast c}$ represents the polarisation vector of gluon $k_i$ where the superscript $c$ is the colour index and the subscript $i$ labels the emitter. Recall that we are assuming energy-ordering for all final-state gluons $\omega_1 \gg \omega_2 \gg \cdots \gg \omega_{m} \gg \omega$, with $\omega_i$ and $\omega$ the energies of gluon $k_i$ and the softest gluon $k$ respectively. 

Iterating the amplitude \eqref{eq:EikAmp_A} down to the Born level, then evaluating the corresponding conjugate amplitude and multiplying out one obtains for the eikonal amplitude squared, of the emission of $m$ soft energy-ordered gluons in a given configuration $X$, the following factorised form \cite{Delenda:Preparaion}:
\begin{subequations}
 \begin{eqnarray}
 W^{\X} \left(p_a,p_b, k_1, \hdots, k_m \right) &=& \mathcal{B}(p_a, p_b) \times \cW^{\X}_{12\cdots m}\,,
\\
 \cW^{\X}_{12\cdots m} &=& \asb^m\; C^{i_1 \cdots i_{m}}_{j_1 \cdots j_{m}}\; \left[\prod_{n=1}^{m} \left(\sum_{i_{n}, j_{n} \in U_{m-1}}\; w^n_{i_{n} j_{n} }   \right) \right]\,,
 \label{eq:EikAmp_B}
\end{eqnarray}
\end{subequations}
where $\mathcal{B}(p_a, p_b)$ is the Born amplitude squared, $\asb = \as/\pi = (g^2_s/4\pi)/\pi$, $U_{m} = \{a,b,1,2,\cdots, m\}$ is the set of all possible emitting dipole-legs of the softest gluon $m$, and the antenna function $w^\ell_{ij}$ is defined by:
\begin{equation}
 w^\ell_{ij} = \omega_\ell^2 \frac{ (h_i \cdot h_j) }{(h_i \cdot h_\ell) (h_j \cdot h_\ell)}\,,
 \label{eq:AntennaFun}
\end{equation}
with $h_a = p_a, h_b = p_b$ and $h_i = k_i$. The colour factor in the above amplitude squared reads:
\begin{equation}
 C^{ i_1 \cdots i_{m}}_{j_1 \cdots j_{m}} = \frac{1}{\mathrm{N}_c}\, \mathrm{tr} \left(\bT^{a_1}_{i_1} \cdots \bT^{a_m}_{i_{m}} \, \bT^{a_1}_{j_1} \cdots \bT^{a_m}_{j_{m}} \right) \,,
 \label{eq:EikColor}
\end{equation}
where ``tr'' means the trace. The colour factor \eqref{eq:EikColor} involves the term $\delta_{q\bar{q}} = \mathrm{tr}\left( \mathbf{1}\right) =$ N$_c$, which is the colour factor associated with the Born amplitude squared. We have absorbed a factor N$_c$ into $\mathcal{B}(p_a, p_b)$ and divided it out in \eqref{eq:EikColor} so as to completely separate the two squared amplitudes. The configuration $\X$ may be written as $\X \mathrm{= x_1 x_2 \cdots x_m}$, with each $\mathrm{x_i \in \{R, V\}}$ where $\mathrm{R}$ stands for real and $\mathrm{V}$ for virtual.

\subsection{Implementation in Mathematica }

The eikonal amplitude squared in eq. \eqref{eq:EikAmp_B} can be computed using the {\tt Mathematica} program ``{\tt EikAmp}'' (see ref. \cite{Delenda:Preparaion}). The main algorithm of the latter is shown in {\bf Algorithm}~\ref{alg:EikAmp}.
\begin{algorithm}[h!]
\caption{Compute eikonal amplitude squared at finite-N$_c$}
\label{alg:EikAmp}
\begin{algorithmic}[1]
\State {\bf Determine} the loop order $m$
\State {\bf Begin} with the bra amplitude:
    \For {each emitted gluon $\ell$}
        \State {\bf Determine} the set of all possible emitting dipole-legs
        \State {\bf Pick} up the first leg of each dipole ($i$)
        \State {\bf Determine} the corresponding colour matrix $\bT^{c_\ell}_i$ (fundamental or adjoint representation)
	
       \State {\bf Move} to the ket amplitude:
          \For {each emitted gluon}
            \State {\bf Determine} the set of all possible emitting dipole-legs 
            \State {\bf Pick} up the second leg of each dipole ($j$)
            \State {\bf Determine} the corresponding conjugate colour matrix $\bT^{c_\ell}_j$
       \EndFor
   
\State {\bf Call} the {\tt CSimplify} function of {\tt ColorMath}
\State {\bf Decompose} the output of {\tt CSimplify} in terms of $\CF = (\mathrm{N}_c^2-1)/2 \mathrm{N}_c$ and $\CA = \mathrm{N}_c$
\State {\bf Multiply} by the appropriate antenna $w^\ell_{ij}$
\EndFor 
\State {\bf Sum} up all contributions (all possible combinations of emitting dipoles)
\end{algorithmic}
\end{algorithm}

The output of the {\tt EikAmp} program is a lengthy tedious expression involving products of colour factors and antenna functions. We use the symmetries learned at 2 and 3 loops to write the resultant expressions at 4 (and partially 5) loops in a closed compact form. The details are to be found in ref. \cite{Delenda:Preparaion}.

\section{Results and discussion}

Based on the results of the {\tt EikAmp} program for 2, 3, 4 and 5 loops the following important properties of the above eikonal amplitude squared, eq. \eqref{eq:EikAmp_B}, may be deduced:
\begin{itemize}
 
 \item It is totally symmetric under the interchange of the two hardest partons; quark and antiquark $(a \leftrightarrow b)$.
 
 \item It is totally symmetric under the interchange of the legs of the dipole emitting the softest gluon.
 
  \item For the softest gluon, it is always true that
 \begin{equation}
  \cW^{\mathrm{x_1 \cdots R}}_{12\cdots m} = - \cW^{\mathrm{x_1 \cdots V}}_{12\cdots m}\,,
 \end{equation}
regardless of the nature, R or V, of the rest of the harder gluons.
 
 \item The first finite-N$_c$ corrections appear at 4 loops. Unlike large-N$_c$ contributions, the former corrections are not symmetric under permutations of gluons. 
  
  \item It seems, contrary to the findings of ref. \cite{Dokshitzer:1991wu}, that the 4 loops amplitude squared has no singular dependence on angles and is fully integrable over the directions of all four gluons involved (in ref. \cite{Khelifa-Kerfa:2015mma} we integrated it out for the hemisphere mass distribution and found a finite answer).
  
 \item In addition to being non-symmetric under permutations of gluons, the eikonal amplitude squared at 5 loops, and perhaps beyond, is not symmetric under the interchange of the legs of each and every single emitting dipole. This symmetry breaking is primarily due to the associated colour factor given in eq. \eqref{eq:EikColor}.  
 
 \item The source of the symmetry-breaking mentioned above is the finite-N$_c$ corrections. Large-N$_c$ contributions are free of any such symmetry-breaking terms, as may easily be verified within the {\tt EikAmp} program.

 \item At large-N$_c$, the eikonal amplitude squared reduces to the simpler expression: 
 \begin{equation}
  \cW^{\X}_{12\cdots m} = \left( \asb\, \mathrm{N}_c \right)^m \, \sum_{\pi_m} \frac{ (p_a p_b)}{ (p_a k_{1}) (k_1 k_2) \cdots (k_m p_b) }\,,
 \end{equation}
 where $\pi_m$ represents all possible permutations of the set $\{p_a, p_b, k_1, \hdots, k_m \}$. This result can straightforwardly be found by setting $\CF \to \CA/2$ in {\tt EikAmp} at any given loop-order in PT.   

\end{itemize}

\section{Summary}

We have been able to overcome the two hindrances that have long jeopardised progress in QCD eikonal matrix-element calculations at finite-N$_c$ beyond leading (2 loop) order. We have developed a {\tt Mathematica} program that performs such calculations in an automated fashion, and which will be made public in the near future. Alternative versions of the program, written in {\tt c++} and/or {\tt fortran}, will also be considered.
As a next step forward it seems natural to extend the present work to the next-to-eikonal approximation. The latter guarantees the resummation of next-to-single logs (of the form $\alpha_s^n L^{n-1}$), leading thus to more precise calculations.

\acknowledgments
This work is supported in part by CNEPRU Research Project D01320130009.

\end{document}